\documentclass[12pt]{article}

\catcode`\@=11
\@addtoreset{equation}{section}

\global\arraycolsep=2pt
\usepackage{mathrsfs,amsbsy,amssymb,latexsym,amsfonts,amsmath,cite}
\usepackage{graphicx,color}

\newcommand\txt {\textrm}

\allowdisplaybreaks

\begin{document}

\begin{flushright}
\parbox{4cm}
{KUNS-2610}
\end{flushright}

\vspace*{1.5cm}

\begin{center}
{\Large \bf
Generalized quark-antiquark potentials \\
from a $q$-deformed AdS$_5 \times $S$^5$ background}
\vspace*{1.5cm}\\
{\large Takashi Kameyama*$^{,\dagger}$\footnote{E-mail:~
takashi.kameyama@yukawa.kyoto-u.ac.jp}
and Kentaroh Yoshida*\footnote{E-mail:~kyoshida@gauge.scphys.kyoto-u.ac.jp}}
\end{center}

\vspace*{0.25cm}

\begin{center}
{*\it Department of Physics, Kyoto University, \\
Kyoto 606-8502, Japan}
\end{center}
\begin{center}
{$^{\dagger}$\it Yukawa Institute for Theoretical Physics, Kyoto University, \\
Kyoto 606-8502, Japan}
\end{center}

\vspace{2cm}

\begin{abstract}
We study minimal surfaces with a single cusp
in a $q$-deformed AdS$_5\times$S$^5$ background.
The cusp is composed of two half-lines with an arbitrary angle
and is realized on a surface specified in the deformed AdS$_5$.
The classical string solutions attached to this cusp
are regarded as a generalization of configurations studied
by Drukker and Forini in the undeformed case.
By taking an antiparallel-lines limit, a quark-antiquark potential for the $q$-deformed case
is derived with a certain subtraction scheme.
The resulting potential becomes linear at short distances with a finite deformation parameter.
In particular, the linear behavior for the gravity dual of noncommutative gauge theories
can be reproduced as a special scaling limit. Finally we study the near straight-line limit 
of the potential. 
\end{abstract}

\setcounter{footnote}{0}
\setcounter{page}{0}
\thispagestyle{empty}

\newpage

\tableofcontents

\section{Introduction}

One of the most profound subjects in String Theory is the AdS/CFT correspondences \cite{M}.
A prototypical example is the conjectured equivalence
between type IIB string theory on the AdS$_5\times$S$^5$ background and
4D $\mathcal{N} = 4$ $SU(N)$ super Yang-Mills (SYM) theory in the large $N$ limit.
A great discovery is that an integrable structure underlying this duality has been unveiled.
This integrability enables us to compute physical quantities at arbitrary coupling constant
even in non-BPS sectors, and it has led to an enormous amount of support for this duality \cite{review}.

\medskip

The classical action of the AdS$_5\times$S$^5$ superstring can be constructed
by following the Green-Schwarz formulation with a supercoset \cite{MT}:
\begin{eqnarray}
\frac{PSU(2,2|4)}{SO(1,4) \times SO(5)}\,,
\end{eqnarray}
which ensures the classical integrability in the sense of kinematical integrability \cite{BPR}.
As a next step, it would be intriguing to consider integrable deformations of the AdS/CFT
and reveal the fundamental mechanism underlying gauge/gravity dualities
without relying on the conformal symmetry.

\medskip

On the string-theory side, in order to study integrable deformations,
it is nice to follow the Yang-Baxter sigma model approach \cite{Klimcik}.
This is a systematic way to consider integrable deformations of 2D non-linear sigma models.
Following this approach, one can specify an integrable deformation by taking a (skew-symmetric)
classical $r$-matrix satisfying the modified classical Yang-Baxter equation (mCYBE).
The deformed sigma-model action is classically integrable
in the sense of the kinematical integrability (i.e., a Lax pair exists).

\medskip

The original argument was restricted to principal chiral models,
but it has been generalized to the symmetric coset case \cite{DMV}\footnote{
For earlier developments on sigma-model realizations of $q$-deformed $su(2)$ and $sl(2)$\,,
see \cite{KYhybrid,KMY-QAA,KOY,Sch}.}.
With this success, a $q$-deformation of the AdS$_5\times$S$^5$ superstring action
has been studied in \cite{DMV-string} by adopting a classical $r$-matrix
of Drinfeld-Jimbo type \cite{DJ}. This deformed system is often called the $\eta$-model.
The metric and NS-NS two-form are computed in \cite{ABF}.
Some special limits of deformed AdS$_n\times$S$^n$ were studied in\cite{HRT,LRT}.
The supercoset construction was recently performed in \cite{ABF2}
and the full background was derived (for the associated solution, see \cite{HT}).
The resulting background does not satisfy the equations of motion of type IIB supergravity,
but it is conjectured that it should satisfy the modified type IIB supergravity equations \cite{ABF-HRT}.

\medskip

For the $\eta$-model, 
a great deal of work has
been done so far.
A mirror description is proposed in \cite{mirror1,mirror2}.
The fast-moving string limits are considered in \cite{KY-LL}.
Giant magnon solutions are studied in \cite{mirror1,magnon}.
The deformed Neumann-Rosochatius systems are derived in \cite{AM-NR, Kame-coord}.
A possible holographic setup has been proposed in \cite{Kame-coord}
and minimal surfaces are studied in \cite{Kame-coord,Kame-surf,Kame-proceeding,BCW}.
Three-point functions \cite{3pt} and the D1-brane \cite{D1} are also discussed.
For two-parameter deformations, see \cite{HRT,bi-DMV,bi-YB,bi-DLMV}.
Another integrable deformation called the $\lambda$-deformation \cite{Sfetsos,Hollowood}
is closely linked to the $\eta$-model by a Poisson-Lie duality
\cite{Vicedo,Hoare-Tseytlin,Sfetsos,Klimcik-lambda}.

\medskip

A possible generalization of the Yang-Baxter sigma model is based
on the homogeneous classical Yang-Baxter equation (CYBE) \cite{KMY-JordanianAdSxS}\,.
A strong advantage is that partial deformations of AdS$_5\times$S$^5$ can be studied.
In a series of works \cite{KMY-SUGRA, MY1, MY2, Sch-YB, YB1, YB2, Stijn, Lax}\,,
a lot of classical $r$-matrices have been identified with the well-known type IIB supergravity solutions
including the $\gamma$-deformations of S$^5$ \cite{LM},
gravity duals for noncommutative (NC) gauge theories \cite{HI}
and Schr\"odinger spacetimes \cite{10DSch}. The relationship between the gravity solutions
and the classical $r$-matrices is referred to as the gravity/CYBE correspondence
(for a short summary, see\cite{CYBE}). This correspondence indicates that
the moduli space of a certain class of type IIB supergravity solutions may be identified with the CYBE.

\medskip

In the recent, this identification has been generalized to integrable deformations
of 4D Minkowski spacetime in \cite{Minkowski}.
A $q$-deformation of the flat space string
has also been studied from a scaling limit of the $\eta$-deformed  AdS$_5\times$S$^5$ \cite{Stijn-kappa}.
Furthermore, as an application,
the Yang-Baxter invariance of the Nappi-Witten model \cite{NW}
has been shown in \cite{Kyono}. Another remarkable feature is that
Yang-Baxter deformations can be applied to non-integrable backgrounds beyond integrability.
An example of non-integrable backgrounds is AdS$_5\times T^{1,1}$ \cite{KW}
and the non-integrability is supported by the existence of chaotic string solutions \cite{BZ,Penrose-T11}.
Then TsT transformations of $T^{1,1}$ can be reproduced as Yang-Baxter deformations as well \cite{CMY-T11}.
This result indicates that the gravity/CYBE correspondence is not limited to the integrable backgrounds.

\medskip

Here we will focus upon minimal surfaces in the $q$-deformed
AdS$_5 \times $S$^5$ superstring \cite{DMV-string,ABF} (the $\eta$-model).
It is interesting to argue a holographic relation in the $q$-deformed background.
We have proposed that the singularity surface in the deformed AdS
may be treated as the holographic screen \cite{Kame-coord,Kame-surf}.
For this purpose, it is convenient  to introduce a coordinate system which describes only the spacetime
enclosed by the singularity surface \cite{Kame-coord}.
Applying this coordinate system, minimal surfaces whose boundaries are straight lines and circles
have been considered in \cite{Kame-coord,Kame-surf}.
In the $q\to 1$ limit, the solutions correspond to a 1/2 BPS straight Wilson loop
\cite{Wilson,Kawamoto}  and a 1/2 BPS circular Wilson loop \cite{BCFM,DGO,Drukker}, respectively.

\medskip

In this paper, we continue to study minimal surfaces in the $q$-deformed AdS$_5\times$S$^5$
and consider a generalization with a single cusp.
The cusp is composed of two half-lines with an arbitrary angle
and it is realized on a surface specified in the deformed AdS$_5$.
The classical string solutions attached to this cusp
are regarded as a generalization of configurations studied
by Drukker and Forini \cite{DF} in the undeformed case. 
By taking an antiparallel-lines limit, we derive a quark-antiquark potential for the $q$-deformed case
with a certain subtraction scheme.
The resulting potential becomes linear at short distances with finite deformation parameter.
In particular, the linear behavior for the gravity dual for noncommutative gauge theories can be reproduced
by taking a special scaling limit.

\medskip

This paper is organized as follows.
Section 2 gives a short review of string theory on the $q$-deformed AdS$_5 \times $S$^5$ background.
In section 3, we study a minimal surface whose boundary is given by two half-lines with an angle.
The classical action is evaluated with a certain subtraction scheme.
In section 4, a quark-antiquark potential is derived by taking an antiparallel-lines limit.
The resulting potential exhibits a linear behavior at short distances.
In Section 5, we study the near straight-line expansion in detail.
Section 6 is devoted to the conclusion and discussion.

\medskip

Appendix A  summarizes the definition of elliptic integrals and some properties.
Appendix B presents a classical solution of the Wilson loop with a cusp  in the global coordinates.
Appendix C gives a review of the derivation of a linear potential at short distances
from the gravity dual of noncommutative gauge theories.

\section{String theory on a $q$-deformed AdS$_5\times$S$^5$}

The classical superstring action on the $q$-deformed AdS$_5\times$S$^5$ background
has been constructed by Delduc, Magro and Vicedo \cite{DMV-string}.
The metric (in the string frame) and NS-NS two-form have been computed in \cite{ABF}.
Here, for simplicity, we will focus upon  the bosonic part of the classical action
with the conformal gauge. The resulting action is composed of the metric part
and the Wess-Zumino (WZ) term, as we will show later.

\subsection{The convention of the string action \label{sec-action}}

Let us here introduce the bosonic part of the string action
(with the conformal gauge), which can be divided into the metric part $S_G$
and the Wess-Zumino (WZ) term $S_{\rm WZ}$:
\begin{eqnarray}
S &=& S_G + S_{\rm WZ}\,. \nonumber
\end{eqnarray}
Here $S_{\rm WZ}$  describes  the coupling of string to an NS-NS two-form.

\medskip

The metric part $S_G$ is further divided into the deformed AdS$_5$ and S$^5$ parts like
\begin{eqnarray}
S_G &=& \int\!d\tau d\sigma\, \left[\,
\mathcal{L}_G^{\rm (AdS)} + \mathcal{L}_G^{\rm (S)}\,
\right] \nonumber \\
& = &-\frac{1}{4\pi \alpha'}\int\!\!d\tau d\sigma\,\eta^{\mu\nu}
\left[G_{MN}^{\rm (AdS)}\, \partial_{\mu}X^M \partial_{\nu}X^N +
G_{PQ}^{\rm (S)}\, \partial_{\mu}Y^P \partial_{\nu}Y^Q \right]\,, \nonumber
\end{eqnarray}
where the string world-sheet coordinates are $\sigma^{\mu}=(\sigma^0,\sigma^1)=(\tau,\sigma)$
with $\eta_{\mu\nu}=(-1,+1)$\,.

\medskip

The WZ term $S_{\rm WZ}$ is also divided into two parts:
\begin{eqnarray}
S_{\rm WZ} &=& -\frac{1}{4\pi\alpha'}\int\!d\tau d\sigma\, \epsilon^{\mu\nu}\left[\,
B_{MN}^{\rm (AdS)}\, \partial_{\mu}X^M \partial_{\nu}X^N +
B_{PQ}^{\rm (S)}\, \partial_{\mu}Y^P \partial_{\nu}Y^Q \right]\,.\nonumber
\end{eqnarray}
Here the totally anti-symmetric tensor $\epsilon^{\mu\nu}$ is normalized as $\epsilon^{01}=+1$\,.

\subsection{A $q$-deformed AdS$_5\times$S$^5$ \label{sec-ABF}}

Next, let us introduce the metric and NS-NS two-form of a $q$-deformed AdS$_5\times$S$^5$\cite{ABF}\footnote{By performing a supercoset construction,
the full component has been obtained recently \cite{ABF2}.}.

\medskip

The metric is divided into the deformed AdS$_5$ and S$^5$ parts like
\begin{eqnarray}
ds^2_{(\textrm{AdS}_5)_q}&=& R^2\sqrt{1+C^2}\,
\Bigl[ -\dfrac{\cosh^2\rho\,dt^2}{1-C^2\sinh^2\rho}
+\dfrac{d\rho^2}{1-C^2\sinh^2\rho}
+\dfrac{\sinh^2\rho\,d\zeta^2}{1+C^2\sinh^4\rho\sin^2\zeta}\nonumber\\
&&\hspace{2.3cm}+\dfrac{\sinh^2\rho\cos^2\zeta\,d\varphi^2}{1+C^2\sinh^4\rho\sin^2\zeta}
+\sinh^2\rho\sin^2\zeta\,d\psi^2\Bigr]\,,\label{ads}
\\\nonumber\\
ds^2_{(\textrm{S}^5)_q} &=& R^2\sqrt{1+C^2}\,
\Bigl[ \dfrac{\cos^2\gamma\,d\vartheta^2}{1+C^2 \sin^2\gamma}
+\dfrac{d\gamma^2}{1+C^2\sin^2\gamma}
+\dfrac{\sin^2\gamma\,d\xi^2}{1+C^2\sin^4\gamma\sin^2\xi}\nonumber\\
&&\hspace{2.3cm}+\dfrac{\sin^2\gamma\cos^2\xi\,d\phi_1^2}{1+C^2\sin^4\gamma\sin^2\xi}
+\sin^2\gamma\sin^2\xi\,d\phi_2^2\Bigr]\,.
\end{eqnarray}
Here $(t\,, \varphi\,, \psi\,,\zeta\,, \rho)$  parameterize the deformed AdS$_5$\,, while
$(\vartheta\,, \gamma\,, \phi_1\,, \phi_2 \,, \xi)$ does the deformed S$^5$\,.
The deformation is characterized by a real parameter $C \in [0,\infty)$\,.
When $C=0$\,, the geometry is reduced to the usual AdS$_5\times$S$^5$
with the curvature radius $R$\,. Note that  a curvature singularity exists
at $\rho=\txt{arcsinh}\,(1/C)$ in (\ref{ads})\,.

\medskip

Let us comment on the causal structure near the singularity surface \cite{Kame-surf}.
For massless particles, it takes infinite affine time to reach the singularity surface, while it does not
in the coordinate time. Massive particles cannot reach the surface as well.
Thus this property is essentially the same as the conformal boundary of the usual global AdS$_5$\,.

\medskip

The NS-NS two-form $B_2 = B_{(\txt{AdS}_5)_q} + B_{(\txt{S}^5)_q}$ is given by
\begin{eqnarray}
B_{(\txt{AdS}_5)_q} &=& R^2\,C\sqrt{1+C^2}\,\dfrac{\sinh^4\rho\sin2\zeta}{1+C^2\sinh^4\rho\sin^2\zeta}\,
d\varphi \wedge d\zeta\,, \\
B_{(\txt{S}^5)_q}&=&-R^2\,C\sqrt{1+C^2}\,\dfrac{\sin^4\gamma\sin2\xi}{1+C^2\sin^4\gamma\sin^2\xi}\,
d\phi_1  \wedge d\xi\,.
\label{wz}
\end{eqnarray}
Note that $B_2$ vanishes when $C=0$\,.

\subsection{Poincar\'e coordinates}

In Sec.\,\ref{sec-ABF}, we have discussed the bosonic background in the global coordinates \cite{ABF}.
In order to study minimal surfaces, however, it is helpful to adopt Poincar\'e-like coordinates
for the metric of the deformed  AdS$_5$ (\ref{ads}) in the Euclidean signature.

\medskip

After performing  the Wick rotation $t \to i\tau$ and the  coordinate transformation:
\begin{eqnarray}
\sinh\rho=\frac{r}{\sqrt{z^2+C^2(z^2+r^2)}}\,,\qquad \textrm{e}^\tau=\sqrt{z^2+r^2}\,,
\end{eqnarray}
the resulting metric describes a deformed Euclidean AdS$_5$\cite{Kame-surf},
\begin{eqnarray}
ds^{2}_{(\textrm{AdS}_5)_q}&=& R^2\sqrt{1+C^2}\,
\biggl[ \frac{dz^2+dr^2}{z^2+C^2(z^2+r^2)}
+ \frac{C^2(z\,dz+r\,dr)^2}{z^2\bigl(z^2+C^2(z^2+r^2)\bigr)}
\label{poincare} \\
&& + \frac{\bigl(z^2+C^2(z^2+r^2)\bigr)r^2}{
\bigl(z^2+C^2(z^2+r^2)\bigr)^2+C^2r^4\sin^2\zeta}\,(d\zeta^2+\cos^2\zeta\,d\varphi^2 )
+ \frac{r^2\sin^2\zeta\,d\psi^2}{z^2+C^2(z^2+r^2)}\biggr]\,.\nonumber
\end{eqnarray}
Note that the singularity surface is now located at $z=0$ in the above coordinates as well.
It is shown in \cite{Kame-surf} that space-like proper distances to the singularity surface are finite,
in comparison to the undeformed case. This property might be important in the next section.

\medskip

When $C=0$\,, the deformed metric (\ref{poincare}) is reduced to the Euclidean AdS$_5$
with the Poincar\'e coordinates as a matter of course. Inversely speaking, one may think of
the metric (\ref{poincare}) giving rise to an integrable deformation of Euclidean Poincar\'e AdS$_5$
from the beginning.

\section{Cusped minimal surfaces}

In this section, we will explicitly derive a classical string solution ending on two half-lines
with an arbitrary angle (i.e., a cusp) on the boundary of the Euclidean $q$-deformed AdS$_5$\,.
After solving the equations of motion, the Nambu-Goto action is evaluated with a boundary term
coming from a Legendre transformation. Our argument basically follows seminal papers \cite{DGRT,DF}
for the undeformed case.

\subsection{Classical string solutions}

Suppose that the boundary conditions are lines separated
by $\pi-\phi$ on the boundary of the deformed AdS$_5$ and $\theta$ on the deformed S$^5$\,.
Hence it is sufficient to consider an AdS$_3\times$S$^1$ subspace of the deformed background
by imposing that the solution is located at\footnote{It seems quite difficult to study 
a cusped minimal surface solution with a non-vanishing $\zeta$ on the $q$-deformed background, 
hence we choose $\zeta=0$ for simplicity.} 
\begin{eqnarray}
\psi=\zeta=0~~(\mbox{AdS$_5$})\,,\qquad\gamma=\phi_1=\phi_2=\xi=0~~(\mbox{S$^5$})\,.
\label{ads3xs1}
\end{eqnarray}
Then the resulting metric of the AdS$_3\times$S$^1$ subspace is given by
\begin{eqnarray}
ds^{2}_{(\textrm{AdS}_3\times\textrm{S}^1)_q}&=& R^2\sqrt{1+C^2}\,
\biggl[ \frac{dz^2+dr^2+r^2d\varphi^2}{z^2+C^2(z^2+r^2)}
+ \frac{C^2(z\,dz+r\,dr)^2}{z^2\bigl(z^2+C^2(z^2+r^2)\bigr)}\label{metric}
+ d\vartheta^2
\biggr]
\,.\label{ds}
\end{eqnarray}
Note that the NS-NS-two form (\ref{wz})  vanishes under the condition (\ref{ads3xs1})\,.

\medskip

It should be remarked that the metric (\ref{poincare}) (or (\ref{ds})) is invariant under the rescaling
\begin{eqnarray}
z \to c_0\,z\,, \qquad r \to c_0\,r \qquad (c_0 >0)\,. \label{rescaling}
\end{eqnarray}
Then it is helpful to suppose that the tip of the cusp is located at $r = 0$
so that the cusp is invariant under the rescaling (\ref{rescaling}).

\medskip

Let us take $r$ and $\varphi$ as the string world-sheet coordinates.
Then it is natural to suppose that the $z$ coordinate is linear to $r$
so as to respect the rescaling (\ref{rescaling}) like
\begin{eqnarray}
z=r\,v(\varphi)\,,\qquad \vartheta= \vartheta(\varphi)\,.
\label{ansatz}
\end{eqnarray}
The coordinate $\varphi$ extends from $\phi/2$ to $\pi-\phi/2$\,.
We suppose that $v(\varphi)=0$ at the two boundaries where $\varphi=\phi/2$ and $\pi-\phi/2$\,,
while it takes its maximal value $v_m$ at $\varphi=\pi/2$\,.
For the sphere part, the coordinate $\vartheta$ ranges from $-\theta/2$ to $\theta/2$\,.

\medskip

Then the Nambu-Goto action is given by
\begin{eqnarray}
S_{\textrm{NG}}&=&\sqrt{1+C^2}\,\frac{\sqrt{\lambda}}{2\pi}\, \int \frac{dr}{r}
\int_{\phi/2}^{\pi-\phi/2}\!\! d\varphi\,\mathcal{L}(\varphi)\,,\nonumber\\
\mathcal{L}(\varphi)&=&
\frac{1}{v\,v_C}\sqrt{v'^2+(1+v^2)(1+v_C^2\,\vartheta'^2)}\,.
\label{nambu}
\end{eqnarray}
Here the prime is the derivative with respect to $\varphi$\,, and $\lambda$ is  defined as
\begin{eqnarray}
\sqrt{\lambda}\equiv\dfrac{R^{2}}{\alpha'}\,.
\end{eqnarray}
We have introduced the $C$-dependent function $v_C$ as
\begin{eqnarray}
v_C(\varphi)\equiv\sqrt{v^2+C^2(1+v^2)}\,,
\label{vc}
\end{eqnarray}
which is reduced to $v$ in the $C\to0$ limit.

\medskip

Note that the $r$-dependence is factored out, hence two conserved charges can readily be obtained.
The Hamiltonian $E$ which corresponds to  $\partial_\varphi$ translations
and the canonical momentum $J$ conjugate to $\vartheta$ are given by, respectively,
\begin{eqnarray}
E=\frac{1+v^2}{v\,v_C\sqrt{v'^2+(1+v^2)(1+v_C^2\,\vartheta'^2)  }}\,,\qquad
J=\frac{v_C(1+v^2)\,\vartheta'}{v\sqrt{v'^2+(1+v^2)(1+v_C^2\,\vartheta'^2)  }}\,.
\label{EJ}
\end{eqnarray}
It is helpful to introduce the following conserved quantities:
\begin{eqnarray}
p\equiv\frac{1}{E}>0\,,\qquad q\equiv\frac{J}{E}=
v_C^2\,\vartheta'\,.
\label{pq}
\end{eqnarray}
Then, by using the relation (\ref{pq}), the differential equation for $v(\varphi)$ is given by
\begin{eqnarray}
v'^2&=&
\frac{1+v^2}{v^2\,v_C^2}\left[p^2+(p^2-q^2)v^2-v^2v_C^2\right]
\label{eqf} \\
&=& \frac{1+v^2}{v^2\,v_C^2}\frac{p^2}{v_m^2}\left(1+\frac{b^2}{p^2}\,v^2\right)(v_m^2-v^2)\,.
\nonumber
\end{eqnarray}
Here a new parameter $b$ has been introduced  as
\begin{eqnarray}
b \equiv \sqrt{\frac{1}{2}\left(p^2-q^2-C^2+\sqrt{ (p^2-q^2-C^2)^2+4(1+C^2)p^2}\,\right)} >  0
\,,\label{b}
\end{eqnarray}
and $v_m$ is the turning point in $v$ where  $v'(\varphi=\pi/2)=0$\,,
which is given by
\begin{eqnarray}
v_m^2 = \frac{b^2}{1+C^2}\,.
\label{f0}
\end{eqnarray}

\medskip

To make the equation (\ref{eqf}) more tractable, let us define the following function,
\begin{eqnarray}
x(\varphi) \equiv \sqrt{\frac{v_C^2\left(b^4+(1+C^2)p^2\right)}{(1+C^2)(b^2+C^2)(p^2+b^2v^2)}}\,.
\label{x}
\end{eqnarray}
Then the equation (\ref{eqf})  can be expressed as an elliptic equation,
\begin{eqnarray}
x'^2=\frac{b^2\left[b^4+(1+C^2)p^2\right]}{\left[p^2+C^2(p^2-b^2)\right]^2}
\left(\frac{b^4+(1+C^2)p^2}{b^2(b^2+C^2)x^2}-1\right)^2(1-x^2)(1-k^2x^2)\,,
\label{eqx}
\end{eqnarray}
where $k$ is defined as
\begin{eqnarray}
k \equiv \sqrt{\frac{(1+C^2)(b^2+C^2)(b^2-p^2)}{b^4+(1+C^2)p^2}}\,,
\label{def-k}
\end{eqnarray}
and it  satisfies $ 0\leq k<1$\,. Note that $k$ becomes zero when $E=|J|$\,,
i.e., $q=\pm 1$\,, with the condition $p>0\,,C\geq 0$\,.
This corresponds to a BPS case in the undeformed limit.

\medskip

For later purposes, it may be helpful to rewrite $p$ and $q$ in terms of $b$ and $k$ as
\begin{eqnarray}
p^2 &=& \frac{b^2\left[(1+C^2)(b^2+C^2)-b^2k^2\right]}{(1+C^2)(b^2+C^2+k^2)}\,,\\
q^2 &=& \frac{(1+C^2)(b^2+C^2)-k^2\left[b^4+(1+C^2)(2b^2+C^2)\right]}{(1+C^2)(b^2+C^2+k^2)}\,.
\label{pq-bk}
\end{eqnarray}

\medskip

In the $C\rightarrow0$ limit, $b$ and $k$ are reduced to
the undeformed ones $b_0$ and $k_0$ \cite{DF}, respectively,
\begin{eqnarray}
b^2 \rightarrow b_0^2\equiv \frac{1}{2}\left(p^2-q^2+\sqrt{ (p^2-q^2)^2+4p^2}\,\right)\,,
\qquad
k^2 \rightarrow k_0^2\equiv \frac{b_0^2(b_0^2-p^2)}{b_0^4+p^2}\,.
\label{b0}
\end{eqnarray}
Note that $p$ and $q$ above are given by (\ref{EJ}) and (\ref{pq}) with $C=0$\,.

\subsubsection*{Classical solution}

Let us solve the equation of motion for $x(\varphi)$\,.

\medskip

In the first place, one needs to fix a boundary condition.
Here let us impose the following boundary condition.
For the world-sheet segment i) $\phi/2\leq\varphi\leq\pi/2$\,, $v(\varphi)$
increases monotonically as
\begin{eqnarray}
v(\phi/2)=0\,~~(\textrm{boundary})\,~~\longrightarrow\,~~v(\pi/2)=v_m\,
\quad (\textrm{midpoint})\,,\label{bc1}
\end{eqnarray}
while for the other segment ii)  $\pi/2<\varphi\leq\pi-\phi/2$\,, $v(\varphi)$
decreases monotonically as
\begin{eqnarray}
v(\pi/2)=v_m \,~~(\textrm{midpoint})\,~~\longrightarrow\,~~v(\pi-\phi/2)=0\,
\quad (\textrm{boundary})\,.
\end{eqnarray}
In terms of $x(\varphi)$  (\ref{x})\,,
the above conditions can be rewritten as
\begin{eqnarray}
&{\rm i)} & \qquad x(\phi/2)=x_0\,~~(\textrm{boundary})\,~~\longrightarrow\,~~x(\pi/2)=1\,
\quad (\textrm{midpoint})\,,
\nonumber\\
&{\rm ii)}&\qquad x(\pi/2)=1 \,~~(\textrm{midpoint})\,~~\longrightarrow\,~~x(\pi-\phi/2)=x_0\,
\quad (\textrm{boundary})\,.\nonumber
\end{eqnarray}
At the boundary, $x(\varphi)$ takes the minimum value $x_0$ given by
\begin{eqnarray}
x_0=\frac{C}{\sqrt{1+C^2}}\sqrt{\frac{b^4+(1+C^2)p^2}{p^2(b^2+C^2)}}\,.
\label{def-x0}
\end{eqnarray}
Note that $x_0$ satisfies $0<x_0<1$ and vanishes in the $C\rightarrow0$ limit.
At the turning point $\varphi=\pi/2$\,,  $x(\varphi)$ take the maximum, i.e., $x(\pi/2)=1$\,.

\medskip

Then let us solve the first-order differential equation (\ref{eqx})\footnote{Here the $(+)$-signature
is taken, because we consider a classical solution stretching from a boundary at $\varphi=\phi/2$
to the turning point at $\varphi=\pi/2$\,. },
\begin{eqnarray}
x'&=&\frac{b\sqrt{b^4+(1+C^2)p^2}}{p^2+C^2(p^2-b^2)}
\left(\frac{b^4+(1+C^2)p^2}{b^2(b^2+C^2)x^2}-1\right)\sqrt{(1-x^2)(1-k^2x^2)}\,.
\label{eq-x}
\end{eqnarray}
By integrating (\ref{eq-x}) for $\varphi \geq \phi/2$ with the boundary condition $x(\phi/2)=x_0$\,,
one can obtain the following expression:
\begin{eqnarray}
\int_{\phi/2}^\varphi \!\!d\tilde{\varphi}=
\frac{p^2+C^2(p^2-b^2)}{b\sqrt{b^4+(1+C^2)p^2}}\int_{x_0}^x \!\!
\frac{d\tilde{x}}{\left(\frac{b^4+(1+C^2)p^2}{b^2(b^2+C^2)
\tilde{x}^2}-1\right)\sqrt{(1-\tilde{x}^2)(1-k^2\tilde{x}^2)}}\,.
\end{eqnarray}
Then the world-sheet coordinate
$\varphi$ ($\phi/2\leq\varphi\leq\pi/2$) can be written
in terms of incomplete elliptic integrals of the first and third kinds like
\begin{eqnarray}
\varphi&=&\frac{\phi}{2}
+\frac{p^2+C^2(p^2-b^2)}{b\sqrt{b^4+(1+C^2)p^2}}\biggl[\Pi\left(\frac{b^2(b^2+C^2)}{b^4+(1+C^2)p^2},
\textrm{arcsin}\,x\,\Bigl|k^2\right)-F(\textrm{arcsin}\,x\,|k^2)
\nonumber \\&&\hspace{2cm}-\Pi\left(\frac{b^2(b^2+C^2)}{b^4+(1+C^2)p^2},
\textrm{arcsin}\,x_0\,\Bigl|k^2\right)
+F(\textrm{arcsin}\,x_0\,|k^2)\biggr]\,.
\end{eqnarray}
By taking $x(\pi/2)=1$\,, the cusp angle $\pi-\phi$ is represented by
\begin{eqnarray}
\phi&=&\pi-
2\,\frac{p^2+C^2(p^2-b^2)}{b\sqrt{b^4+(1+C^2)p^2}}
\biggl[\Pi\left(\frac{b^2(b^2+C^2)}{b^4+(1+C^2)p^2}\,
\Bigl|k^2\right)-K(k^2) \nonumber \\
&&\hspace{2cm}-\Pi\left(\frac{b^2(b^2+C^2)}{b^4+(1+C^2)p^2},
\textrm{arcsin}\,x_0\,\Bigl|k^2\right)
+F(\textrm{arcsin}\,x_0\,|k^2)\biggr]\,.
\label{phi}
\end{eqnarray}
Note here that the incomplete elliptic integrals have been replaced by complete ones.

\medskip

For the other segment $\pi/2 < \varphi \leq \pi-\phi/2$\,,
an analytical continuation of the solution is necessary.
The resulting expression is given by
\begin{eqnarray}
\varphi&=&\pi-\frac{\phi}{2}
-\frac{p^2+C^2(p^2-b^2)}{b\sqrt{b^4+(1+C^2)p^2}}\biggl[\Pi\left(\frac{b^2(b^2+C^2)}{b^4+(1+C^2)p^2},
\textrm{arcsin}\,x\,\Bigl|k^2\right)-F(\textrm{arcsin}\,x\,|k^2)
\nonumber \\&&\hspace{2cm}-\Pi\left(\frac{b^2(b^2+C^2)}{b^4+(1+C^2)p^2},\textrm{arcsin}\,x_0\,\Bigl|k^2\right)
+F(\textrm{arcsin}\,x_0\,|k^2)\biggr]\,.
\label{x-sol2}
\end{eqnarray}
By taking the $C\to0$ limit with $p$ and $q$ fixed,
the solutions (\ref{phi}) and  (\ref{x-sol2}) are reduced to the undeformed ones \cite{DF}.

\subsubsection*{The sphere part}

The remaining thing is to get the expression of $\vartheta(\varphi)$\,.
First of all, let us fix a boundary condition.
For the world-sheet segment i)  $\phi/2\leq\varphi\leq\pi/2$\,,
$\vartheta(\varphi)$ increases monotonically as
\begin{eqnarray}
\vartheta(\phi/2)=-\theta/2\,~~(\textrm{boundary})\,
~~\longrightarrow\,~~\vartheta(\pi/2)=0\,
~~(\textrm{midpoint})\,,
\end{eqnarray}
while for the other segment ii) $\pi/2<\varphi\leq\pi-\phi/2$\,,
$\vartheta(\varphi)$ decreases monotonically as
\begin{eqnarray}
\vartheta(\pi/2)=0 \,~~(\textrm{midpoint})\,
~~\longrightarrow\,~~\vartheta(\pi-\phi/2)=\theta/2\,
~~(\textrm{boundary})\,.
\end{eqnarray}
By integrating $\vartheta$ in (\ref{pq}) for  $\varphi \geq  \phi/2$\,,
one can obtain the following expression: 
\begin{eqnarray}
\int_{-\theta/2}^\vartheta\!\! d\tilde{\vartheta}=\frac{b\,q}{\sqrt{b^4+(1+C^2)p^2}}
\int_{x_0}^{x}\frac{d\tilde{x}}{\sqrt{(1-\tilde{x}^2)(1-k^2\tilde{x}^2)}}\,.
\end{eqnarray}
This equation can be understood as the relation between $x$ and $\vartheta$\,,
\begin{eqnarray}
\vartheta=-\frac{\theta}{2}+\frac{b\,q}{\sqrt{b^4+(1+C^2)p^2}}\left[F(\textrm{arcsin}\,x\,|k^2)
-F(\textrm{arcsin}\,x_0\,|k^2)
\right]\,.
\label{theta-sol}
\end{eqnarray}
When $\varphi=\pi/2$\,, i.e., $x(\pi/2)=1$\,, it reaches the midpoint $\vartheta=0$\,,
hence the angle $\theta$ in the sphere part is determined as
\begin{eqnarray}
\theta=\frac{2b\,q}{\sqrt{b^4+(1+C^2)p^2}}\left[K(k^2)
-F(\textrm{arcsin}\,x_0\,|k^2)
\right]\,.\label{theta}
\end{eqnarray}
As a result, the expression (\ref{theta-sol}) can be rewritten as
\begin{eqnarray}
\vartheta=\frac{b\,q}{\sqrt{b^4+(1+C^2)p^2}}\left[F(\textrm{arcsin}\,x\,|k^2)-K(k^2)
\right]\,.
\end{eqnarray}
This solution can also reproduce the result in \cite{DF} in the undeformed limit.

\subsection{The classical action}

The next step is to evaluate the value of the classical string action.

\medskip

First of all, it is helpful to rewrite the Euclidean classical action (\ref{nambu})
by using the equation of motion (\ref{eqx})\,. The resulting action is given by
\begin{eqnarray}
S_{\textrm{NG}}=
\frac{\sqrt{\lambda}}{2\pi}\,\frac{\sqrt{1+C^2}}{C}\,\frac{x_0\sqrt{1-x_0^2}}{\sqrt{1-k^2x_0^2}}\,
\int^{R_{\textrm{IR}}}_{\epsilon_{\textrm{UV}}}
\frac{dr}{r}\, 2\int_{x_0}^{1}\!\!dx\,\frac{\sqrt{1-k^2x^2}}{(x^2-x_0^2)\sqrt{1-x^2}}\,.
\end{eqnarray}
Here $R_{\textrm{IR}}$ and $\epsilon_{\textrm{UV}}$ are IR and UV cut-offs
for the $r$-direction, respectively.
Then the $r$-integral is evaluated as
\begin{eqnarray}
\int^{R_{\textrm{IR}}}_{\epsilon_{\textrm{UV}}}
\frac{dr}{r}=\log\frac{R_{\textrm{IR}}}{\epsilon_{\textrm{UV}}} \equiv T\,.\label{cutoffT}
\end{eqnarray}
Note that this quantity $T$ can be identified with
an interval of the global time coordinate\footnote{The $r$ coordinate
is identified with the global time $t$ through $r=\txt{exp}\,t$\,,
hence $\int dr/r=\int \! dt\equiv T$\,.}.
Then the classical action can be calculated as
\begin{eqnarray}
S_{\textrm{NG}}&=&
-\frac{T\sqrt{\lambda}}{\pi}\,
\frac{\sqrt{1+C^2}}{C}\,\frac{x_0\sqrt{1-x_0^2}}{\sqrt{1-k^2x_0^2}}
\left[k^2\int_{x_0}^{1}\frac{dx}{\sqrt{1-x^2}\sqrt{1-k^2x^2}} \right.
\nonumber\\
&&\hspace{5cm} \left. +(x_0^{-2}-k^2)
\int_{x_0}^{1}\frac{dx}{(1-x_0^{-2}x^2)\sqrt{1-x^2}\sqrt{1-k^2x^2}}
\,\right] \nonumber \\
&=&-\frac{T\sqrt{\lambda}}{\pi}\,
\frac{\sqrt{1+C^2}}{C}\,\frac{x_0\sqrt{1-x_0^2}}{\sqrt{1-k^2x_0^2}}\,
\biggl(k^2\left[K\left(k^2\right)-F\bigl(\textrm{arcsin}\,x_0
\,|k^2\bigr)\right]  \nonumber \\
&&\hspace{2cm}+\left(x_0^{-2}-k^2\right) \Bigl[ \Pi\left(x_0^{-2}\,|k^2\right)-
\lim_{\epsilon\to0}\Pi\bigl(x_0^{-2}\,,
\textrm{arcsin}(x_0+\epsilon)\,|k^2\bigr)\Bigr]\biggr)\,.
\label{S-NG}
\end{eqnarray}
where $\Pi(\alpha^2,\psi|k^2)$ is an incomplete elliptic integral of the third kind (for details, see Appendix A).
In general, $\Pi(\alpha^2\,, \psi\,|k^2)$ are interpreted as Cauchy principal values
when $\alpha>1$ \cite{Elliptic}\,, and there are singularities
on the real axis at $\psi=\txt{arcsin}\,\alpha^{-1}$\,.
Thus, a cut-off $\epsilon$ has been introduced as $\epsilon \equiv x-x_0$ for the limit $x\to x_0$\,,
because $\Pi\bigl(x_0^{-2}\,,\textrm{arcsin}\,x\,|k^2\bigr)$ diverges logarithmically as $x\to x_0$\,.
Through the relation (\ref{x})\,,
the cut-off $\epsilon$ can be converted into $v_0$\,, which is a cut-off for small $v$:
\begin{eqnarray}
\epsilon\equiv
x-x_0 = \frac{x_0(1-k^2x_0^2)}{2C^2}\,v_0^2
+ \mathcal{O}(v_0^{4})\,.
\label{epsilon}
\end{eqnarray}

\medskip

The elliptic integrals in (\ref{S-NG}) can be rewritten as follows:
\begin{eqnarray}
&& \Pi\left(x_0^{-2}\,|k^2\right)-
\lim_{\epsilon\to0}\Pi\bigl(x_0^{-2}\,,
\textrm{arcsin}(x_0+\epsilon)
\,|k^2\bigr)\nonumber\\
&=&\lim_{\epsilon\to0}\left(\int_0^{x_0-\epsilon}dx+\int_{x_0+\epsilon}^{1}dx\right)
\frac{1}{(1-x_0^{-2}x^2)\sqrt{1-x^2}\sqrt{1-k^2x^2}}\nonumber\\&&\hspace{2cm}-
\lim_{\epsilon\to0}\int_0^{x_0-\epsilon}dx\,\,\frac{1}{(1-x_0^{-2}x^2)\sqrt{1-x^2}\sqrt{1-k^2x^2}}
\nonumber\\
&=&PV\,\Pi\left(x_0^{-2}|k^2\right)-\lim_{\epsilon\to 0}
\Pi\bigl(x_0^{-2}\,,
\textrm{arcsin}(x_0-\epsilon)
\,|k^2\bigr)\,.
\nonumber
\end{eqnarray}
Then the principal value can be evaluated as\footnote{Use the formula 415.01 in \cite{Elliptic}.}
\begin{eqnarray}
PV\,\Pi\left(x_0^{-2}|k^2\right) =
-\frac{x_0\,K\left(k^2\right)\,Z\left(\textrm{arcsin}\,x_0|k^2\right)}{\sqrt{(1-x_0^2)(1-k^2x_0^2)}}\,,
\end{eqnarray}
where $Z\left(\psi\,|k^2\right)$ is a Jacobi Zeta function given by
\begin{eqnarray}
Z\left(
\psi\,|k^2\right)&\equiv&E\left(
\psi\,|k^2\right)-\frac{E(k^2)}{K(k^2)}\,F\left(
\psi\,|k^2\right)\,.
\end{eqnarray}
The incomplete elliptic integral of the third kind can be rewritten
as\footnote{Use the formula 436.01 in \cite{Elliptic}.}
\begin{eqnarray}
\Pi\bigl(x_0^{-2}\,, \textrm{arcsin}(x_0-\epsilon)\,|k^2\bigr)
&=&
\frac{x_0}{\sqrt{(1-x_0^2)(1-k^2x_0^2)}}\biggl(\frac{1}{2}\,\log\Bigl[
\frac{\vartheta_1(\omega+\nu,q_k)}{\vartheta_1(\omega-\nu,q_k)}\Bigr]
\nonumber\\&&\hspace{0cm}-F\bigl(\textrm{arcsin}\,(x_0-\epsilon)
\,|k^2\bigr)\,Z\left(\textrm{arcsin}\,x_0|k^2\right)\biggr)\,,
\label{pi}
\end{eqnarray}
where $\vartheta_1(z,q_k)$ is the Jacobi theta function.
The parameters $\nu\,,\omega$ and $q_k$ are defined as
\begin{eqnarray}
\nu \equiv \frac{\pi F\bigl(\textrm{arcsin}(x_0-\epsilon)
\,|k^2\bigr)}{2K(k^2)}\,,\quad
\omega \equiv \frac{\pi F\left(\textrm{arcsin}\,x_0\,|k^2\right)}{2K(k^2)}\,,
\quad q_k \equiv \textrm{e}^{-\frac{\pi K(1-k^2)}{K(k^2)}}\,.
\end{eqnarray}

\subsection{Separation of the divergence}

Let us decompose $S_{\txt{NG}}$ (\ref{S-NG}) into the finite part $S_{\rm ren}$
and the divergent part $S_{0}$ like
\begin{eqnarray}
S_{\txt{NG}}=S_{\txt{ren}}+S_0\,.
\label{divide}
\end{eqnarray}
Here $S_{\txt{ren}}$ and $S_0$ are given by, respectively,
\begin{eqnarray}
S_{\txt{ren}} &=& \frac{T\sqrt{\lambda}}{\pi}\,\frac{\sqrt{1+C^2}}{C}\,
\biggl(K\left(k^2\right)\,Z\left(\textrm{arcsin}\,x_0|k^2\right)
\label{S-reg} \\
&&\hspace{4cm} -\frac{k^2x_0\sqrt{1-x_0^2}}{\sqrt{1-k^2x_0^2}}
\left[K\left(k^2\right)-F\bigl(\textrm{arcsin}\,x_0\,|k^2\bigr)\right]
\biggr)\,,\nonumber\\\nonumber\\
S_0 &=& \lim_{\epsilon\to 0}
\frac{T\sqrt{\lambda}}{2\pi}\,
\frac{\sqrt{1+C^2}}{C}\,\log\left[\frac{\vartheta_1(\omega+\nu,q_k)}{\vartheta_1(\omega-\nu,q_k)}\,
\txt{e}^{-2\,F\left(\textrm{arcsin}(x_0-\epsilon)\,|k^2\right)\,Z\left(\textrm{arcsin}\,x_0\,|k^2\right)}\right]\,.
\label{S0}
\end{eqnarray}

\medskip

It is useful to look the origin of the logarithmic divergence in (\ref{S0}) in more detail.
Let us convert the $\epsilon$-dependence
to the $v_0$-one through the relation (\ref{epsilon}).
Then the divergent part $S_0$ can be expanded around $v_0 =0$ as
\begin{eqnarray}
S_0 &=& \frac{T\sqrt{\lambda}}{\pi}\,\frac{\sqrt{1+C^2}}{C}\,\Biggl(
\log\biggl[\frac{2C}{\sqrt{1+C^2}\,v_0}\biggr]
\label{s0-se}\\
&& +\log\biggl[\Bigl(\frac{(1+C^2)\sqrt{1-x_0^2}\, K\left(k^2\right)\,
\vartheta_1(2\omega\,,q_k)}{\pi x_0\sqrt{1-k^2x_0^2}\,
\vartheta'_1(0\,,q_k)}\Bigr)^{\frac{1}{2}}\txt{e}^{-\,
F\left(\textrm{arcsin}\,x_0\,|k^2\right)\,Z\left(\textrm{arcsin}\,x_0\,|k^2\right)}
\biggr]\Biggr) +\mathcal{O}(v_0^{2})\,,\nonumber
\end{eqnarray}
where $\vartheta'_1(\omega\,,q_k)=\partial_\omega\vartheta_1(\omega\,,q_k)$\,.
Now one can see that the first log-term in (\ref{s0-se}) diverges logarithmically as $v_0\to0$\,,
while the second log is finite. It is worth noticing that the second log-term vanishes
in the undeformed limit $C\to0$\,.

\medskip

Here we should remark that there is an ambiguity that the second log-term in (\ref{s0-se})
may be included in $S_{\rm ren}$\,. In particular, the second log-term vanishes
in the undeformed limit $C\to0$\,, and hence one cannot remove this ambiguity
by relying on the undeformed limit. Therefore,  consistency
with the undeformed limit is not enough, and it is necessary to adopt an extra criterion.

\medskip

Fortunately, there is a definite answer to this issue, i.e.,
 a scaling limit of the $q$-deformed AdS$_5\times$S$^5$
to the gravity dual for NC gauge theories \cite{ABF2}.
Consistency with this limit give rises to a sufficiently strong constraint for the regularization.
In summary, we will adopt the following criteria to regularize the Nambu-Goto action:
\begin{enumerate}
\item[a)~~] $S_{\txt{ren}}$ is reduced to the usual regularized action in the $C\to0 $ limit.
\item[b)~~] The antiparallel-lines limit of $S_{\txt{ren}}$ reproduces
a quark-antiquark potential derived from the gravity dual for NC gauge theories
by taking the scaling limit \cite{ABF2}.
\end{enumerate}

\medskip

According to these criteria, the second log-term in (\ref{s0-se}) should NOT be included in $S_{\rm ren}$
so as to satisfy condition $b)$\,, as we will see later.
Thus, what is the physical interpretation of the second log-term?
In the next subsection, we will consider the physical interpretation of the regularization.

\subsection{Interpretation of the regularization}

We will consider here the physical interpretation of the regularization adopted in the previous subsection.
Our aim is to compute the quark-antiquark potential and hence it is necessary to subtract the contribution
of the static quark mass from the Nambu-Goto action. In addition, we have to take account of
the Legendre transformation as usual. In the following, we will see the contributions of the quark mass
and the Legendre transformation. Finally, we will check the consistency with the undeformed limit.

\subsubsection*{The quark mass}

Let us first evaluate the static  quark mass in the present case.
The total mass of a quark and an antiquark $\widetilde{S}_0$
is given by two strings which  stretch between the boundary ($v=0$)
and the origin of the deformed AdS$_5$ ($v=\infty$)\,, with a constant of $\varphi$:
\begin{eqnarray}
\widetilde{S}_0&=&2\,\frac{T\sqrt{\lambda}}{2\pi}\, \sqrt{1+C^2}\,\int_{\widetilde{v}_0}^{\infty}\,
\frac{dv}{v\,v_C}
\nonumber\\
&=&\frac{T\sqrt{\lambda}}{\pi}\,\frac{\sqrt{1+C^2}}{C}\, \txt{arcsinh}
\Bigl[\frac{C}{\sqrt{1+C^2}\,\widetilde{v}_0}\Bigr]\,.
\label{mass}
\end{eqnarray}
Here a cut-off $\widetilde{v}_0~(\ll 1)$ has been introduced for small $v$\,.
When $C$ is fixed, $\widetilde{S}_0$ can be expanded with respect to $\widetilde{v}_0$ like
\begin{eqnarray}
\widetilde{S}_0=
\frac{T\sqrt{\lambda}}{\pi}\,\frac{\sqrt{1+C^2}}{C}\,
\log\left[\frac{2C
}{\sqrt{1+C^2}\,\widetilde{v_0}}
\right]+\mathcal{O}(v_0^{2})
\,.
\label{mass2}
\end{eqnarray}

\medskip

Now one can identify $\widetilde{S}_0$ with $S_0$ in (\ref{s0-se})
through the following relation:
\begin{eqnarray}
\widetilde{v}_0\equiv v_0\,\biggl(\frac{\pi x_0\sqrt{1-k^2x_0^2}\,
\vartheta'_1(0\,,q_k)}{(1+C^2)\sqrt{1-x_0^2}\, K\left(k^2\right)\,
\vartheta_1(2\omega\,,q_k)}\biggr)^{\frac{1}{2}}
\txt{e}^{F\left(\textrm{arcsin}\,x_0\,|k^2\right)\,
Z\left(\textrm{arcsin}\,x_0\,|k^2\right)}\,.
\label{v0}
\end{eqnarray}
This relation (\ref{v0}) can also be expanded around $C=0$ like
\begin{eqnarray}
\widetilde{v}_0= v_0 +\mathcal{O}(C^2)\,,
\end{eqnarray}
and hence $\widetilde{v}_0$ is equal to $v_0$ in the undeformed limit.
In other words, there is a slight difference between $\widetilde{v}_0$ and $v_0$
in the deformed case, and it should be interpreted as a renormalization effect.

\subsubsection*{Legendre transformation}

The next step is to examine an additional contribution
which comes from the boundary condition \cite{DGO}\,.
The total derivative term $S_L$ is given by
\begin{eqnarray}
S_L= \int\! dr\,2\! \int_{\phi/2}^{\pi/2}\!\!d\varphi\,\partial_\varphi
\Bigl(z\,\frac{\partial L}{\partial(\partial_{\varphi} z)}\Bigr)\,.
\end{eqnarray}
By using (\ref{eq-x})\,,  $S_L$ is evaluated as
\begin{eqnarray}
S_L
&=& \sqrt{1+C^2}\,\frac{\sqrt{\lambda}}{2\pi}\int
\frac{dr}{r}\, \frac{2\sqrt{b^4+(1+C^2)p^2}}{b\,p}\!\int_{x_0}^{1}\!\!dx\,\partial_x\Bigl(
\frac{1}{x}\sqrt{\frac{1-x^2}{1-k^2x^2}}\,\Bigr) \nonumber \\
&=& -\sqrt{1+C^2}\,\frac{T\sqrt{\lambda}}{2\pi} \frac{2\sqrt{b^4+(1+C^2)p^2}}{b\,p\,x_0}
\sqrt{\frac{1-x_0^2}{1-k^2x_0^2}}\,.
\label{SL}
\end{eqnarray}
Here the $r$-integral has led to $\int dr/r=T$ as in (\ref{cutoffT})\,.
Note that the expression of (\ref{SL}) is just a constant term,
\begin{eqnarray}
S_L
=-\frac{T\sqrt{\lambda}}{\pi}\,\frac{\sqrt{1+C^2}}{C}\,,
\label{sl}
\end{eqnarray}
while it diverges as $C\to0$\,.

\medskip

This constant term is necessary to add to the Nambu-Goto action
so as to ensure the undeformed limit, as we will see below.

\subsubsection*{The undeformed limit}

Finally, we shall consider the undeformed limit.

\medskip

By expanding $S_{\txt{ren}}$ in (\ref{S-reg}) around $C=0$\,,
the result in \cite{DF} can be reproduced as
\begin{eqnarray}
S_{\txt{ren}} = -\frac{T\sqrt{\lambda}}{\pi}\,
\frac{\sqrt{1+b_0^2}}{b_0}\,\frac{E\left(k_0^2\right)-(1-k_0^2)K
\left(k_0^2\right)}{\sqrt{1-k_0^2}}\,.
\label{Sc0}
\end{eqnarray}
Here $b_0$ and $k_0$ are defined as the $C \to 0$ limit
of $b$ and $k$ in (\ref{b0})\,, respectively.

\medskip

The remaining step is to consider the undeformed limit of $\widetilde{S}_0 (=S_0)$\,.
In the $C \to 0$ limit, $\widetilde{S}_0$ in (\ref{mass2}) has to be canceled out
with $S_L$ in (\ref{sl})\,.
The sum of $\widetilde{S} + S_L$ is evaluated as
\begin{eqnarray}
\widetilde{S}_0+S_L=\frac{T\sqrt{\lambda}}{\pi}\,
\frac{\sqrt{1+C^2}}{C}\,\left(\log\Bigl[\frac{2C
}{\sqrt{1+C^2}\,\widetilde{v_0}}
\Bigr]-1\right)\,.
\end{eqnarray}
Thus this expression tells us that, for the consistency,
the undeformed limit should be taken
as the following double scaling limit:
\begin{eqnarray}
C\to0\,,\qquad\widetilde{v_0}\to0\qquad\txt{with}\qquad
\log\left[\frac{2C}{\widetilde{v_0}} \right]=1\,:\txt{fixed}\,.
\end{eqnarray}

\medskip

In the next section, we will derive a quark-antiquark potential
from the regularized action $S_{\txt{ren}}$
by taking the antiparallel-lines limit of the cusped configuration.

\section{A quark-antiquark potential\label{sec-potential}}

In this section, we will derive a quark-antiquark potential for the $q$-deformed case.

\medskip

It seems quite difficult to realize a rectangle
as the boundary of a string solution
because the boundary geometry is deformed
and hence the rectangular shape is not respected.
On the other hand, a quark-antiquark potential can be evaluated
from the cusped minimal surface solution by taking an antiparallel-lines limit
as in the undeformed case \cite{DF}, though the potential is valid only
at short distances by construction.
The anti-parallel lines limit is realized by taking $\phi\to\pi$\,,
and then a quark-antiquark potential is obtained
as a function of $\pi-\phi\to L$\,.

\medskip

In the undeformed case, the antiparallel-lines limit leads to
a Coulomb potential of $-1/L$
as expected from the conformal symmetry of the $\mathcal{N}=4$ SYM.
However, the conformal symmetry is broken in the $q$-deformed case
(though the scaling invariance survives the deformation).
Hence the resulting potential may be more complicated at short distances,
while it should still have a Coulomb form at large distances because
the IR region of the deformed geometry
is the same as the usual AdS$_5$\,.

\medskip

In the following, we first examine the antiparallel-lines limit with the finite-$C$ case
and derive the the potential.
Then we shall consider the limit after expanding around $C=0$
and see how the deformation modifies the Coulomb potential
at short distances $L\ll1$\,.
Our argument here basically follows the analysis in the undeformed case \cite{DF}.

\subsection{A linear potential at short distances}

Let us first  express the classical solution in terms of $ b$ and $k$ instead of $p$ and $q$\,,
by using the algebraic relations in (\ref{pq-bk})\,. In the following, the deformation parameter $C$
is kept finite.

\medskip

Then let us consider the following limit:
\begin{eqnarray}
b\to0\,,\qquad k\,:\,\textrm{fixed}\,,
\label{anti-limit}
\end{eqnarray}
and we will ignore $\mathcal{O}(b^2)$ terms.
This limit is nothing but the antiparallel-lines limit \cite{DF}.

\medskip

By taking the limit (\ref{anti-limit})\,, $\phi$ in (\ref{phi}) approaches $\pi$ as
\begin{eqnarray}
\pi-\phi=2b\,\frac{\sqrt{k^2+C^2}}{1+C^2}\,.
\end{eqnarray}
For the sphere part, $\theta$ in (\ref{theta}) approaches zero as
\begin{eqnarray}
\theta=2b\,\frac{\sqrt{1-k^2}}{C(1+C^2)}\,.
\end{eqnarray}
The regularized action (\ref{S-reg}) is reduced to
\begin{eqnarray}
S_{\txt{ren}}=
\frac{T\sqrt{\lambda}}{\pi}\,\frac{b}{C^2}
\left[\textrm{E}\left(k^2\right)-(1-k^2)\textrm{K}\left(k^2\right)\right]\,.
\end{eqnarray}
Substituting $\pi-\phi$ for $b$\,, the regularized action leads to the following potential,
\begin{eqnarray}
S_{\txt{ren}}=\frac{T\sqrt{\lambda}}{2\pi}\,\frac{1+C^2}{C^2}\,
\frac{\textrm{E}\left(k^2\right)-(1-k^2)\textrm{K}\left(k^2\right)}{\sqrt{k^2+C^2}}\,(\pi-\phi)\,.
\label{potential}
\end{eqnarray}
This potential is linear, unlike the Coulomb potential in the undeformed case.
In particular, the coefficient of $\pi-\phi$ is positive definite,
hence it can be regarded as a string tension of the potential.
This result is quite similar to the potential obtained
from the gravity dual of NC gauge theories \cite{Kitazawa}.
The linear behavior in (\ref{potential}) is consistent
with the potential for NC gauge theories, as we will see below.

\subsection*{A consistent limit to a NC background \label{sec-MR-potential}}

It is worth presenting a connection between the potential (\ref{potential})
and another potential for a gravity dual of NC gauge theories.
The latter result was originally obtained in \cite{Kitazawa}.
To be comprehensive, the derivation of the potential
is given in Appendix \ref{sec-MR}.

\medskip

First of all, we will introduce a scaling limit  \cite{ABF2}
from the $q$-deformed AdS$_5$ to a gravity dual of NC gauge theories.
This limit is realized by rescaling the coordinates like
\begin{eqnarray}
&&z=\txt{exp}\left[\sqrt{C}\,t\right]\frac{\sqrt{C}}{u}\,,\qquad
r=\txt{exp}\left[\sqrt{C}\,t\right]\,,\qquad
\varphi=\frac{\sqrt{C}\,x_2}{\sqrt{1-\mu^2}}\,,\nonumber\\
&&\psi=\frac{\sqrt{C}\,x_1}{\mu}\,,\qquad
\zeta=\txt{arcsin}\,\mu+\sqrt{C}\,x_3\,,
\label{scaling}
\end{eqnarray}
and taking a $C \to 0$ limit.
Then the metric of the  (Euclidean) $q$-deformed  AdS$_5$
with the Poincar\'e coordinates is reduced
to that of a gravity dual of NC gauge theories
\cite{HI}\footnote{Now the deformed S$^5$ part is reduced to the round S$^5$\,,
though the scaling limit is not described here (for details, see \cite{ABF2}).}:
\begin{eqnarray}
ds_{\txt{NC}}^2&=&R^2\left[\frac{du^2}{u^2}+u^2\left(dt^2+dx_1^2+\frac{dx_2^2+dx_3^2}{1+\mu^2\,u^4}\right)\right]\,,\\
B_{\txt{NC}}&=&R^2\,\mu\,\frac{dx_2\wedge dx_3}{1+\mu^2\,u^4}\,.
\end{eqnarray}
This result indicates that the $q$-deformed geometry contains a noncommutative space as a special limit.

\medskip

The next issue is to apply the scaling limit to our classical solution. Let us see the $\zeta$-dependence
of the reduction ansatz (\ref{ads3xs1}) and the scaling limit (\ref{scaling})\,.
In the ansatz (\ref{ads3xs1})\,, $\zeta$ is set to be zero, while $\zeta$ is expanded
around a non-zero constant $\arcsin\mu$\,. To take account of this gap,
it is necessary to take a $\mu \to 0$ limit with a scaling limit (\ref{scaling})\,.

\medskip

Then, after taking the rescaling (\ref{scaling})\,,
the cusped ansatz (\ref{ansatz}) can be rewritten as
\begin{eqnarray}
&& v(\varphi)\equiv\frac{z}{r}=\frac{\sqrt{C}}{u(\sigma)}\,, \qquad
r = \txt{exp}\left[\sqrt{C}\,\tau\right]\,,\qquad
\varphi = \frac{\sqrt{C}\,\sigma}{\sqrt{1-\mu^2}}\,.
\label{scaling2}
\end{eqnarray}
Now $\partial_\varphi v$ is converted to $\partial_\sigma u$ through
\begin{eqnarray}
\partial_\varphi v(\varphi)=\frac{d\sigma}{d\varphi} \,\partial_\sigma \left(\frac{\sqrt{C}}{u(\sigma)}\right)
= -\sqrt{1-\mu^2}\,\frac{\partial_\sigma u(\sigma)}{u(\sigma)^2}\,.
\end{eqnarray}
With the rescaled variables in (\ref{scaling2})\,, $T$ and $\pi-\phi$ are redefined as new parameters
$\widetilde{T}$ and $\widetilde{L}$\,, respectively:
\begin{eqnarray}
T &=& \int \frac{dr}{r}=\int_{-\widetilde{T}/2}^{\widetilde{T}/2} \sqrt{C}\, d\tau
= \sqrt{C}\,\widetilde{T}\,,\nonumber\\
\pi-\phi &=& \int\! d\varphi=\int_{-\widetilde{L}/2}^{\widetilde{L}/2}\frac{\sqrt{C}\,d\sigma}{\sqrt{1-\mu^2}}
= \frac{\sqrt{C}\,\widetilde{L}}{\sqrt{1-\mu^2}}\,.
\label{TL}
\end{eqnarray}
By the use of the rescaling (\ref{scaling2}) and the relations in (\ref{TL})\,,
an antiparallel-lines limit of the regularized action for the $q$-deformed case
can be rewritten as
\begin{eqnarray}
S_{\textrm{NG}}=\frac{\widetilde{T}\sqrt{\lambda}}{2\pi}\,\frac{1+C^2}{C\sqrt{1-\mu^2}}\,
\frac{\textrm{E}\left(k^2\right)-(1-k^2)\textrm{K}\left(k^2\right)}{\sqrt{k^2+C^2}}\,
\widetilde{L}\,.
\label{4.12}
\end{eqnarray}
Then we take a double scaling limit:
\begin{eqnarray}
C \to 0 \quad \& \quad \mu \to 0 \quad \txt{with}\quad \frac{C}{\mu}
=\frac{\sqrt{2}}{k}\left[\textrm{E}\left(k^2\right)-(1-k^2)\textrm{K}
\left(k^2\right) \right] \quad \txt{fixed}\,.
\label{c-mu}
\end{eqnarray}
As a result, the potential (\ref{4.12}) is reduced to
that for the gravity dual of NC gauge theories
(with $\mu\to0$)\,,
\begin{eqnarray}
S_{\textrm{NG}}=\frac{\widetilde{T}\sqrt{\lambda}}{2\pi}\,
\frac{\widetilde{L}}{\sqrt{2}\,\mu}\,.
\label{MR-potential}
\end{eqnarray}
Thus we have checked that the scaling limit (\ref{scaling}) is consistent with
our subtraction scheme.

\subsection{Expansion around $C=0$}

The next step is to study the potential behavior when $C$ is very small.
The classical action $S_{\txt{ren}}$ is first around $C=0$ while $b$ and $k$ are fixed.
Then it is expanded around $b=0$ with $k$ fixed\footnote{
Note that there is an ambiguity in the order of limits and the order is sensitive to the potential behavior. 
The opposite order leads to the expansion of (\ref{potential}) around $C=0$\,, hence the linear behavior remains.
}.

\medskip

As a result, $\phi$ in (\ref{phi}) approaches $\pi$ like
\begin{eqnarray}
\pi-\phi=\frac{2b}{k}\,\left[E\left(k^2\right)-(1-k^2)K\left(k^2\right)\right]+\mathcal{O}((C,b)^2)\,.
\end{eqnarray}
For the sphere part, $\theta$  (\ref{theta}) is expanded as
\begin{eqnarray}
\theta=2\sqrt{1-2k^2}\,K\left(k^2\right)
-\frac{2C\sqrt{1-2k^2}}{b\sqrt{1-k^2}}
+\mathcal{O}((C\,,b)^2)\,.
\end{eqnarray}
Thus the regularized action $S_{\txt{ren}}$ (\ref{S-reg}) results in
\begin{eqnarray}
S_{\txt{ren}}&=&\frac{T\sqrt{\lambda}}{\pi}\,
\biggl[\frac{E\left(k^2\right)-(1-k^2)K\left(k^2\right)}{\sqrt{1-k^2}}
\left(-\frac{1}{b}-\frac{b}{2}\right)
+\frac{C\,k^2}{1-k^2}\left(\frac{1}{b^2}+1\right)\biggr]
+\mathcal{O}((C\,,b)^2)\,.\nonumber
\end{eqnarray}
Substituting $\pi-\phi$ for $b$\,,
the leading terms of $S_{\txt{ren}}$ are evaluated as
\begin{eqnarray}
S_{\txt{ren}}=\frac{T\sqrt{\lambda}}{4\pi}\,
\frac{\left(E\left(k^2\right)-(1-k^2)K\left(k^2\right)\right)^2}{k\sqrt{1-k^2}}\left[-\frac{8}{\pi-\phi}
+\frac{16\,C\,k}{(\pi-\phi)^2\,\sqrt{1-k^2}}\right]\,.
\label{S-qq}
\end{eqnarray}
Note that the first term is a Coulomb-form potential which agrees
with the undeformed result obtained in\cite{DF}\,,
while the second term gives rise to a repulsive force with non-vanishing $C$\,.

\medskip

It is worth noting that the sphere-part contribution vanishes
when $k^2=1/2$\,, i.e., $\theta=0$\,.
Then one can obtain the following expression:
\begin{eqnarray}
S_{\txt{ren}}=\frac{T\sqrt{\lambda}}{4\pi}\,\frac{16\pi^3}{\Gamma\left(\frac{1}{4}\right)^4}
\left[-\frac{1}{(\pi-\phi)}+\frac{2C}{(\pi-\phi)^2}\right]\,.
\label{V-qq}
\end{eqnarray}
The first term of (\ref{V-qq}) precisely agrees with the results of \cite{Wilson} by replacing  $\pi-\phi\to L$\,,
and the second term produces a repulsive force, in comparison to the undeformed case.

\section{Near straight-line expansion}

In the undeformed case, the near straight-line limit  is realized as $\phi\to0$\,. 
In this limit, the cusp disappears and the Wilson loop becomes 
an infinite straight line in R$^4$\,, or a pair of antipodal lines on R$\times$S$^3$\,.

\medskip

Let us study here an analogue of the near straight-line limit in the deformed case
by expanding the classical action around $\phi=\theta=0$\,. 
Now $\phi$ and $\theta$ are expressed in terms of 
the parameters $p$ and $q$\,, and the relevant limit indicates that $p$ becomes large.
Note that the modulus $k$ of the elliptic integrals vanishes as $p\to\infty$\,.
Hence we should first expand the elliptic integrals with small $k$\,, 
and then expand around  $p\gg1$\,.

\medskip

The classical solution is expanded as
\begin{eqnarray}
\phi=\frac{2}{p}\left(\textrm{arccot}\,C+C\right)+\mathcal{O}(p^{-3})\,,\qquad
\theta=\frac{2q}{p}\,\textrm{arccot}\,C+\mathcal{O}(p^{-3})\,. 
\label{phi-theta}
\end{eqnarray}
Note here that the $C\to 0$ limit of (\ref{phi-theta}) can reproduce the undeformed result \cite{DF}
\begin{eqnarray}
\phi=\frac{\pi}{p}+\mathcal{O}(p^{-3})\,,\qquad
\theta=\frac{\pi q}{p}+\mathcal{O}(p^{-3})\,.
\end{eqnarray}
Then the regularized action $S_{\txt{ren}}$ can be expanded as
\begin{eqnarray}
S_{\txt{ren}}&=&\frac{T\sqrt{(1+C^2)\lambda}}{4\pi}\,
\frac{q^2-1}{p^2}\left(4\,\txt{arccot}\,C-\pi\right)+\mathcal{O}(p^{-4})\,.
\label{S-str}
\end{eqnarray}

\medskip 

Although the action (\ref{S-str}) is expressed in terms of $p$ and $q$\,, 
it can be rewritten in terms of $\phi$ and $\theta$ through the relations in (\ref{phi-theta})\,.
The resulting formula is given by
\begin{eqnarray}
S_{\txt{ren}} &=& \frac{T\sqrt{(1+C^2)\lambda}}{4\pi}\,\left(\txt{arccot}\,C-\frac{\pi}{4}\right)  \nonumber \\ 
&& \qquad \times \left[\frac{ \theta^2}{(\textrm{arccot}\,C)^2}-\frac{\phi^2}{(\textrm{arccot}\,C+C)^2}\right]
+\mathcal{O}\bigl((\phi^2,\theta^2)^2\bigr)\,. 
\label{TBA}
\end{eqnarray}
In the $C\to 0$ limit, the undeformed result \cite{DF} is reproduced like
\begin{eqnarray}
S_{\txt{ren}}=\frac{T\sqrt{\lambda}}{4\pi}
\,\frac{\theta^2-\phi^2}{\pi}
+\mathcal{O}\bigl((\phi^2,\theta^2)^2\bigr)\,.
\end{eqnarray}
It would be interesting to try to reproduce the result (\ref{TBA}) from a $q$-deformed Bethe ansatz 
by generalizing the methods for the undeformed case \cite{Correa-TBA,Drukker-TBA}.

\section{Conclusion and discussion}

In this paper, we have studied minimal surfaces with a single cusp
in the $q$-deformed AdS$_5\times$S$^5$ background.
By taking an antiparallel-lines limit, a quark-antiquark potential has been computed
by adopting a certain regularization. The UV geometry is modified due to the deformation
and hence the short-distance behavior may be modified from the Coulomb potential,
while the potential should still be of Coulomb type.
In fact, the resulting potential exhibits linear behavior at short distances with finite $C$\,.
In particular, linear behavior for the gravity dual for noncommutative gauge theories can be reproduced
by taking a scaling limit \cite{ABF2}. Finally we have studied the near straight-line limit of the potential.  

\medskip

The most intriguing problem is to unveil the gauge-theory dual for the $q$-deformed AdS$_5\times$S$^5$\,.
In the undeformed case, the potential behaviors at strong and weak coupling
can be reproduced from the Bethe ansatz \cite{Correa-TBA,Drukker-TBA}.
Namely, the gravity and gauge-theory sides are bridged by the Bethe ansatz. 
In particular, an all-loop expression for the near BPS expansion 
of the quark-antiquark potential on an S$^3$\,, which was obtained in \cite{Correa}, 
has been reproduced by an analytic solution of the TBA \cite{Gromov-TBA}. 
Furthermore, for arbitrary $\phi$ and $\theta$\,, it is shown in \cite{Zoltan} that 
the weak-coupling expansion of the TBA reproduces the gauge-theory result up to two loops. 
Recently, the quantum spectral curve technique \cite{QSC} has been applied to study 
the behavior of quark-antiquark potentials \cite{Gromov}. 
It would be possible to adopt these methods for the deformed case as well, 
possibly via a quantum deformed Bethe ansatz. 

\medskip

We hope that the gauge-theory dual can be revealed in light of our linear potential
and the quantum deformed Bethe ansatz.

\subsection*{Acknowledgments}

We are grateful to Valentina Forini, Ben Hoare, Hikaru Kawai, Takuya Matsumoto
and Stijn van Tongeren for useful discussions.
The work of T.K. was supported by the Japan Society for the Promotion of Science (JSPS).
The work of K.Y. is supported by the Supporting Program for Interaction-based Initiative Team Studies
(SPIRITS) from Kyoto University and by the JSPS Grant-in-Aid for Scientific Research (C) No. 15K05051.
This work is also supported in part by the JSPS Japan-Russia Research Cooperative Program
and the JSPS Japan-Hungary Research Cooperative Program.
This work was supported in part by MEXT KAKENHI No. 15H05888.

\section*{Appendix}

\appendix

\section{Elliptic integrals}

The incomplete elliptic integrals of the first, second and third kinds are given by
\begin{eqnarray}
F\left(\psi\,|k^2\right)&=&\int_0^{\sin\psi }\frac{d x}{\sqrt{(1-x^2)
(1-k^2\sin^2\psi)}}\,,\nonumber\\
E\left(\psi\,|k^2\right)&=&\int_0^{\sin\psi }\frac{dx\,\sqrt{1-k^2\sin^2\psi}}{\sqrt{1-x^2}}\,,
\\
\Pi\left(\alpha^2\,,\psi\,|k^2\right)
&=&\int_0^{\sin\psi}\frac{dx}{(1-\alpha^2x^2)\sqrt{1-x^2}\sqrt{1-k^2x^2}}\,.\nonumber
\end{eqnarray}
When $\psi=\pi/2$\,, these expressions become the complete elliptic integrals,
\begin{eqnarray}
K\left(k^2\right)=F\left(\pi/2\,|k^2\right)\,,\quad
E\left(k^2\right)=E\left(\pi/2\,|k^2\right)\,,\quad
\Pi\left(\alpha^2\,|k^2\right)=\Pi\left(\alpha^2\,,\pi/2\,|k^2\right)\,.\nonumber
\end{eqnarray}
Note that $\Pi\left(\alpha^2\,,\psi\,|k^2\right)$
has a pole at $\psi=\textrm{arcsin}\,\alpha^{-1}$\,.
Thus it should be interpreted as the Cauchy principal value when $\alpha^2 \sin^2\psi>1$\cite{Elliptic}\,.

\section{Classical solutions in global coordinates}

Let us consider here cusped solutions in the AdS$_3\times$S$^1$ geometry
with global coordinates (in the Lorentzian signature).
The global coordinate system for the deformed geometry was originally introduced in \cite{ABF}.
For our aim of studying minimal surfaces, it would be rather helpful to adopt another coordinate system
in which the singularity surface is located at the boundary \cite{Kame-coord}.
Then the deformed AdS$_3\times$S$^1$ geometry is written as
\begin{eqnarray}
ds^{2}_{\txt{AdS}_3\times\txt{S}^1} &=& R^2\sqrt{1+C^2}\,
\left[-\cosh^2\chi\, dt^2+\frac{d\chi^2+\sinh^2\chi \,d\varphi^2}{1+C^2\cosh^2\chi}
+ d\vartheta^2 \right]\,.
\label{global}
\end{eqnarray}
The world-sheet coordinates $\tau$ and $\sigma$ are identified with $t$ and $\varphi$\,, respectively.
The other coordinates are supposed to take the following form:
\begin{eqnarray}
\chi=\chi(\varphi)\,,\qquad\vartheta=\vartheta(\varphi)\,, \qquad
\varphi \in \left[\frac{\phi}{2}\,,\pi-\frac{\phi}{2} \right]\,.
\end{eqnarray}
Note here that $\chi$ diverges at $\varphi=\phi/2$ and $\varphi=\pi-\phi/2$\,.
The minimum of $\chi$ is realized at the middle point $\varphi=\pi/2$\,.
The range of $\vartheta$ is bounded from $-\vartheta/2$ to $\vartheta/2$\,.

\medskip

Then the Nambu-Goto action is given by
\begin{eqnarray}
S_{\txt{NG}}=\sqrt{1+C^2}\,\frac{\sqrt{\lambda}}{2\pi}\,\int\! dt \,d\varphi\,
\cosh\,\chi\,\sqrt{\frac{(\partial_\varphi\chi)^2+\sinh^2\chi}{1+C^2\cosh^2\chi}
+(\partial_\varphi\vartheta)^2}\,.
\label{L-global}
\end{eqnarray}
The energy and the canonical momentum conjugate to $\varphi$
are given by, respectively,
\begin{eqnarray}
E&=&\frac{\sinh^2\chi\,\cosh\chi}{(1+C^2\cosh^2\chi)\sqrt{\frac{(\partial_\varphi\chi)^2
+\sinh^2\chi}{1+C^2\cosh^2\chi}+(\partial_\varphi\vartheta)^2}}\,,\nonumber\\
J&=&\frac{\partial_\varphi\vartheta\,\cosh\chi}{\sqrt{\frac{(\partial_\varphi\chi)^2
+\sinh^2\chi}{1+C^2\cosh^2\chi}+(\partial_\varphi\vartheta)^2}}\,.
\label{global-EJ}
\end{eqnarray}
It is convenient to introduce the following quantities like in the Poincar\'e case,
\begin{eqnarray}
p \equiv \frac{1}{E}\,,\qquad
q \equiv \frac{J}{E}=\frac{1+C^2\cosh^2\chi}{\sinh^2\chi}\,\partial_\varphi\vartheta\,.
\end{eqnarray}
By removing $\partial_{\varphi}\vartheta$\,, one can obtain the following relation:
\begin{eqnarray}
p=\frac{(1+C^2\cosh^2\chi)\left((\partial_\varphi\chi)^2+\sinh^2\chi \right)
+q^2\,\sinh^4\chi}{\sinh^2\chi\,\cosh\chi}\,.
\end{eqnarray}
Then this expression can be rewritten as
\begin{eqnarray}
(\partial_\varphi\chi)^2=\frac{(p^2\,\cosh^2\chi-q^2)\sin^4\chi}{1+C^2\cosh^2\chi}-\sinh^2\chi\,.
\label{eq-chi}
\end{eqnarray}

\medskip

Note here that the resulting expression (\ref{eq-chi}) is the same as (\ref{eqf})
in the Poincar\'e case through the identification
\begin{eqnarray}
\sinh\chi(\varphi) \quad \longleftrightarrow  \quad \frac{1}{v(\varphi)}\,.
\end{eqnarray}
Then, the $t$-dependence of (\ref{L-global}) is translated to the $r$-dependence
of (\ref{nambu}) via $t \,\leftrightarrow\, \log r$\,.

\section{A linear potential at short distances in NC gauge theories\label{sec-MR}}

Let us consider a minimal surface solution in the gravity dual of NC gauge theories \cite{HI}.
This solution is dual to a rectangular Wilson loop on the gauge-theory side.
From the classical action of this solution, a quark-antiquark potential can be evaluated.
Then the resulting potential exhibits a linear behavior at short distances \cite{Kitazawa}.
The following argument is just a short review of \cite{Kitazawa}.

\medskip

The (Euclidean) metric and $B$-field for the gravity dual \cite{HI} are given by
\begin{eqnarray}
ds_{\txt{NC}}^2&=&R^2\left[\frac{du^2}{u^2}+u^2
\left(dt^2+(dx^1)^2+\frac{(dx^2)^2+(dx^3)^2}{1+\mu^2\,u^4}\right)\right]\,, \\
B_{\txt{NC}}&=&R^2\,\mu\,\frac{dx^2\wedge dx^3}{1+\mu^2u^4}\,.
\label{MR}
\end{eqnarray}
Here a constant parameter $\mu$ measures the noncommutative deformation\footnote{
On the gauge-theory side, the $x^2$-$x^3$ plane is deformed to a noncommutative plane
with the noncommutativity $\theta_0$\,. The parameter $\mu$ is related to $\theta_0$
through the following relation \cite{Kitazawa}~:
\begin{eqnarray}
\mu=\sqrt{\lambda}\,\theta_0\,.
\end{eqnarray}
Now that $\lambda$ is assumed to be large for the validity of the gravity dual,
$\mu\gg\theta_0$\,.
}.

\medskip

To derive a quark-antiquark potential from this background,
we study the following configuration of a static string described by
\begin{eqnarray}
u=u(\sigma)\,,\qquad t=\tau\,,\qquad x^2=\sigma\,,\qquad x^1=x^3=0\,.
\label{MR-ansatz}
\end{eqnarray}
Then the classical string describes a rectangular loop on the boundary.

\medskip

Hereafter, we will consider a 4D slice of the metric (\ref{MR})
at $u=\Lambda$ by following \cite{Kitazawa}.
Suppose that $\widetilde{L}$ is the distance between two antiparallel lines at $u=\Lambda$\,,
and $\widetilde{T}$ is an interval for the $\tau$-direction.
As a result, the classical action is rewritten as
\begin{eqnarray}
S_{\textrm{NG}}=\frac{\sqrt{\lambda}}{2\pi}\,\int_{-\widetilde{T}/2}^{\widetilde{T}/2}
d\tau \int_{-\widetilde{L}/2}^{\widetilde{L}/2}\!\! d\sigma \,\sqrt{(\partial_\sigma u)^2+\frac{u^2}{1+\mu^2u^4}
}\,.
\label{NG-MR}
\end{eqnarray}
From this action, the classical solution can be obtained as
\begin{eqnarray}
\frac{\widetilde{L}}{2} &=& \frac{1}{u_m}\int_1^{\Lambda/u_m} \!\!dy\,\frac{1+\mu^2u_m^4\,y^4}{y^2\sqrt{y^4-1}}\,,
\label{L-MR}
\end{eqnarray}
and then the value of the classical action is evaluated as
\begin{eqnarray}
S_{\textrm{NG}} &=& \frac{\widetilde{T}\sqrt{\lambda}}{\pi}\,\sqrt{1+\mu^2u_m^4}\,u_m
\int_1^{\Lambda/u_m}\!\!dy\,\frac{y^2}{\sqrt{y^4-1}}\,.
\label{S-MR}
\end{eqnarray}
Here $u_m$ is the turning point along the $u$-direction where $\partial_\sigma u=0$\,.

\medskip

In the following, we will focus upon a special case, $u_m \sim \Lambda$
and consider the behavior of the classical action (\ref{S-MR}) with a double scaling limit:
\begin{eqnarray}
\mu \to 0\,, \quad \Lambda \to \infty \quad
\mbox{with}~~ \sqrt{\mu}\,\Lambda \equiv 1\,.
\label{Lambda}
\end{eqnarray}
Then the integrands in (\ref{L-MR}) and (\ref{S-MR}) can be expanded in terms of $y$\,.
Hence (\ref{L-MR}) and (\ref{S-MR}) are evaluated as
\begin{eqnarray}
\frac{\widetilde{L}}{2}&=&\frac{1+\mu^2u_m^4}{u_m}\,\sqrt{\frac{\Lambda}{u_m}-1}
+\mathcal{O}\left(\frac{\Lambda}{u_m}-1\right)^{3/2}\,, \\
S_{\textrm{NG}}&=&\frac{\widetilde{T}\sqrt{\lambda}}{\pi}\,\sqrt{1+\mu^2u_m^4}\,u_m\,
\sqrt{\frac{\Lambda}{u_m}-1}+\mathcal{O}\left(\frac{\Lambda}{u_m}-1\right)^{3/2}\,.
\end{eqnarray}
Note here that
the distance $\widetilde{L}$ is very small (in comparison to $\sqrt{\mu}$),
because
\begin{eqnarray}
\frac{\widetilde{L}^2}{\mu} \sim
\left(\frac{\Lambda}{u_m}-1\right)\ll 1\,.
\end{eqnarray}
As a result, $S_{\rm NG}$ can be expressed as a function of $\widetilde{L}$ like
\begin{eqnarray}
S_{\rm NG} &=& \frac{\widetilde{T}\sqrt{\lambda}}{2\pi}\,\widetilde{L}\,\frac{u_m^2}{\sqrt{1+\mu^2u_m^4}}
+ \mathcal{O}\Bigl((\widetilde{L}/\sqrt{\mu})^3\Bigr) \nonumber \\
&\simeq& \frac{\widetilde{T}\sqrt{\lambda}}{2\pi}\,\frac{\widetilde{L}}{\sqrt{2}\,\mu}
+ \mathcal{O}\Bigl((\widetilde{L}/\sqrt{\mu})^3\Bigr)\,. \label{linear-kitazawa}
\end{eqnarray}
Here the relation $\sqrt{\mu}\,u_m \sim1$ has been utilized in the last line.
This expression (\ref{linear-kitazawa}) leads to a linear potential at short distances\footnote{
The usual Coulomb potential can be reproduced 
at long distances.\cite{Kitazawa}.} \cite{Kitazawa}.

\end{document}